\documentclass[conference]{IEEEtran}
\usepackage{multirow}
\usepackage{rotating}
\usepackage{hyperref}
\usepackage{latexsym}
\usepackage{rotating}
\usepackage{amsfonts}
\newif\ifremarks
\remarksfalse 

\def\EDB{{\textsf EDB}}
\def\NVD{{\textsf NVD}}
\def\EKITS{{\textsf EKITS}}
\def\SYM{{\textsf SYM}}
\def\SEDB{{\textsf EDB'}}
\def\SNVD{{\textsf NVD'}}
\def\SEKITS{{\textsf EKITS'}}

\begin{document}

\title{My Software has a Vulnerability, should I Worry?
}

\author{\IEEEauthorblockN{Luca Allodi and Fabio Massacci}
\IEEEauthorblockA{DISI - University of Trento\\
Trento, Italy\\
Email: lastname@disi.unitn.it}
}

\maketitle
\begin{abstract}
(U.S) Rule-based policies to mitigate software risk suggest to use the CVSS
score to measure the individual vulnerability risk and act accordingly: an HIGH
CVSS score according to the \NVD\ (National (U.S.) Vulnerability Database) is
therefore translated into a ``Yes''. A key issue is whether such rule is
economically sensible, in particular if reported vulnerabilities have been
\emph{actually exploited in the wild}, and whether the risk score do actually
match the risk of actual exploitation.

We compare the \NVD\ dataset with two additional datasets, the \EDB\ for the
white market of vulnerabilities (such as those present in Metasploit), and the
\EKITS\ for the exploits traded in the black market. We benchmark them against
Symantec's threat explorer dataset (\SYM) of actual exploit in the wild. We
analyze the whole spectrum of CVSS submetrics and use these characteristics to
perform a \emph{case-controlled analysis} of CVSS scores (similar to those used
to link lung cancer and smoking) to test its reliability as a risk factor for
actual exploitation.

We conclude that (a) fixing just because a high CVSS score in \NVD\ only yields
negligible risk reduction, (b) the additional existence of proof of concepts
exploits (e.g. in \EDB) may yield some additional but not large risk reduction,
(c) fixing in response to presence in black markets yields the equivalent risk
reduction of wearing safety belt in cars (you might also die but still\ldots).
On the negative side, our study shows that as industry we miss a metric with
high specificity (ruling out vulns for which we shouldn't worry).

[In order to address the feedback from BlackHat 2013's audience, the final
revision (V3) provides additional data in Appendix \ref{appendix:controls}
detailing how the control variables in the study affect
the results.]
\end{abstract}



%
\IEEEpeerreviewmaketitle

\section{Introduction}

Software vulnerabilities assessments usually rely on the National (US)
Vulnerability Database\footnote{\url{http://nvd.nist.gov}} (\NVD\ for short).
Each vulnerability is published with its ``technical assessment'' given by the
Common Vulnerability Scoring System\footnote{\url{http://www.first.org/cvss}}
(CVSS) which rate diverse aspects of the vulnerability \cite{Mell-2007-CMU}.

Despite not being designed to be a metric for risk, the CVSS score is often used as
such. For example, the US Federal government with QTA0-08-HC-B-0003
reference notice
specified that IT products to manage and assess the security of IT
configurations must use the NIST certified S-CAP protocol
\cite{Scarfone-2010-SCAP}, which explicitly says: ``\emph{Organizations should
use CVSS base scores to assist in prioritizing the remediation of known
security-related software flaws based on the relative severity of the flaws.}'' In other words,
a rule-based policy is enforced: if the vulnerability is marked as ''high risk'' by the CVSS assessment, it must be fixed with high priority.

This interest from the industry is matched by many academic studies. On one side,
Vulnerability Discovery Models \cite{Alhazmi-2008-TR,Massacci-2012-ASIACCS} try
to predict the number of vulnerabilities that affect a software at a certain
point in time, while empirical studies try to identify trends between open and
closed source software \cite{Frei-2006-LSAD,Shahzad-2012-ICSE}. On the other,
attack graphs \cite{Wang-2008-DAS} and attack surfaces \cite{Howard-2005-CS21C}
aim at assessing in which ways a system is ``attackable'' by an adversary and
how easily he/she can succeed. Foundational to both  approaches is calculating
a) the number of vulnerabilities in the system and b) their individual ``risk
assessment''.

Beside \NVD, many datasets are used in vulnerability studies, but are they the
right databases? For example, Bozorgi et al. \cite{Bozorgi-2010-SIGKDD} showed
(as a side result) that the exploitability CVSS subscore distribution do not
correlate well with existence of known exploit from the ExploitDB. There are
two ways to interpret this result: the exploitability of CVSS is the wrong
metric, or Bozorgi and his co-authors used the wrong DB. ExploitDB could just
be used by security researchers to show off their skills (and obtain more
contracts as penetration testers) but might not have a correlation with actual
attacks by hackers. The same problem is faced in
\cite{Shahzad-2012-ICSE} where a large majority of
``exploits'' are reported as zero-days\footnote{A zero-day exploit is present when the
exploit is reported before or on the date that the vulnerability is
disclosed.}. The ``exploit'' time in OVSDB only measures the time when a
proof-of-concept exploit becomes known. Security researchers
normally submit proof-of-concept exploits to vendors and vulnerability white
markets in order to prove that the vulnerability is worth the
bounty~\cite{Miller-2007-WEIS}. So it is not surprising that there are a lot of
``zero-day'' exploits; still, this does not mean that a bad hacker really exploited those
vulnerabilities.

\subsection{Our Contribution}
In this work we address the following questions:
\begin{enumerate}
\item To what extent can public vulnerability datasets be used to measure software security?
\item Are the rule-based policies enforced, for example,
by the US Government effective in decreasing risk of attacks?
\end{enumerate}
In other words, when new vulnerabilities are found, are we measuring the rate
at which security researchers try to extract bounties from vendors (and should
not worry)? or there is a concrete risk that bad guys end up exploiting our
systems (and should worry)? This is particularly interesting for the majority of
\emph{internet users at large} (individuals or corporations) who have not
enough individual value to justify a targeted attack\footnote{Obviously, for a
nuclear power plan any proof-of-concept exploit is a problem as even a software
crash may lead to a national emergency.}. To this aim we analyzed three
datasets:
\begin{itemize}
\item \NVD, the benchmark universe of vulnerabilities;
\item \EDB\ (Exploit-DB), which contains information on the existence of
    proof-of-concept exploits;
\item \EKITS, our database containing vulnerabilities used in exploit kits
    sold in the black market.
\end{itemize}
No previous study, to the best of our knowledge, extensively looked at CVSS
\emph{sub}scores throughout different datasets. We benchmark these DBs against
the vulnerabilities exploited in the wild that we collected from Symantec's
Threats and Attack Signatures databases (\SYM). To make statistically
sound conclusions, we perform a
case-controlled randomized experiment in which we build random samples of the \NVD,
\EDB\ and \EKITS\ datasets according to the characteristics
of exploits in \SYM; our goal is to
understand the conditional probability that a CVSS (sub)score would lead to an
attack.

The conclusion of our analysis is the following: the \NVD\ and \EDB\ databases
are not a reliable source of information for exploits in the wild, and the CVSS
score doesn't help. The CVSS score only shows a significant sensitivity (i.e.
prediction of attacks in the wild) for vulnerabilities bundled in exploit kits
in the black market (\EKITS). Unfortunately, it does not show a high
specificity in any of our datasets, which means that it is not suitable, as a metric,
to rule out ``un-interesting'' vulnerabilities (i.e. those not exploited).
This leads to a very low (3\%) expected reduction in risk
one would benefit from by using solely the CVSS score as a prioritisation metric
for patching. Including the presence of a proof-of-concept exploit or of an exploit
traded in the black markets can raise the bar up to 60\%.

The fact that \EKITS\ vulnerabilities are actually exploited in the wild is
interesting in its own sake.  ``Malware sales'' are often scams for wanna-be
scammers, such as credit-card numbers sold over IRC channels
\cite{Herley-2010-EISP}. Surprisingly, while the final products (card numbers)
sold on the black market are bad, the software tools to get them from the
source look good.

In the rest of the paper we introduce our four datasets (\S\ref{sec:datasets})
and draws a first, observational comparison (\S\ref{sec:comparison}). The core
of the paper analyses the goodness of the CVSS global score as a test for
exploitation (\S\ref{sec:sensi-speci}), digs down over the submetrics
(\S\ref{sec:exploits-characteristics}),  and identifies trade-offs in the
exploitation (\S\ref{sec:exploitation-tradeoff}). Then we describe our
randomized case-controlled analysis (\S\ref{sec:randomized}), and
 discuss the implication of our findings (\S\ref{sec:discussion}) and
threats to validity (\S\ref{sec:threats}). We finally discuss related works
(\S\ref{sec:related}) and conclude (\S\ref{sec:conclusions}).
For completeness and replicability, we
also report in Appendix \ref{appendix:controls} the whole set of results
from the controlled experiment.

\section{Datasets} \label{sec:datasets}

\NVD\ is the reference database for disclosed vulnerabilities held by NIST. It
has been widely used and analyzed in previous vulnerability studies
\cite{fabioViet:after-life2011,Shahzad-2012-ICSE,Scarfone:2009}. Our \NVD\
dataset contains data on 49599 vulnerabilities.

The Exploit-db\footnote{http://www.exploit-db.com/} (\EDB) includes information
on proof-of-concept exploits also represented in the Open Source Vulnerability
Database (OSVDB). Both
OSVDB\footnote{\url{http://blog.osvdb.org/2012/08/15/august-2012-a-few-small-updates}}
and \EDB\footnote{\url{http://www.exploit-db.com/author/?a=3211&pg=1}} derive
data from Metasploit Framework. \EDB\ references exploited CVEs by each entry
in the db. Most notable studies relying on either EDB or OSVDB are
\cite{Shahzad-2012-ICSE,Bozorgi-2010-SIGKDD}. \EDB\ contains data on 8122
vulnerabilities for which a \emph{proof-of-concept code} is reported.

\EKITS\ is our dataset of vulnerabilities bundled in Exploit
Kits\footnote{Exploit Kits are web sites that the attacker deploys on some
public webserver he/she owns. When the victim is fooled in making an HTTP
connection to the Exploit Kit, the latter checks for vulnerabilities on the
user's system and, if any, tries to exploit them; eventually, it infects the
victim machine with malware of some sort.} sold on the black market.
\EKITS\ is a substantial expansion on Contagio's Exploit Pack Table\footnote{\url{http://contagiodump.blogspot.it/2010/06/overview-of-exploit-packs-update.html}}.
\EKITS\ repots exploits for 114 unique CVEs bundled in 90+ exploit kits. We cannot disclose the
individual sources of the black-hat communities because this might hamper us
from future studies.

In order to determine whether a vulnerability has been used in the wild we have
collected exploited CVEs from Symantec's
AttackSignature\footnote{\url{http://www.symantec.com/security\_response/attacksignatures/}}
and
ThreatExplorer\footnote{\url{http://www.symantec.com/security\_response/threatexplorer/}}
public data. The \SYM\ dataset contains 1277 CVEs identified in viruses
(local threats) and remote attacks (network threats) by Symantec's commercial
products. This has of course
some limitation as direct attacks by individual motivated hackers against
specific companies are not considered in this metric. 
Note that \SYM\ does not report volumes of exploits, but only the binary information
``evidence of an exploit in the wild for that CVE is reported'' or ``is not reported''.

Table \ref{tab:datasets} summarizes the content of each dataset and the
collection methodology. They are available upon request\footnote{\url{http://securitylab.disi.unitn.it/doku.php?id=datasets}}.
\begin{table}[t]
\centering
\caption{Summary of our datasets} \label{tab:datasets}
\begin{tabular}{|p{1cm} | p{2cm} | p{3cm} | c|}
\hline\hline
DB & Content & Collection method & \#Entries\\
\hline\hline
NVD & CVEs & XML parsing & 49599\\\hline
EDB & Publicly exploited CVEs & Download and web parsing to correlate with CVEs & 8122\\\hline
SYM & CVEs exploited in the wild & Web parsing to correlate with CVEs & 1277 \\\hline
EKITS & CVEs in the black market & ad-hoc analysis + Contagio's Exploit table &114 \\
\hline\hline
\end{tabular}
\end{table}

\section{Exploratory analysis of datasets}\label{sec:comparison}

As a starting point, we perform an exploratory analysis of our four datasets:
\emph{Given a dataset (\NVD, \EDB, \EKITS), what is the likelihood that a
vulnerability it contains is going to be exploited in the wild?} i.e. occurs
also in \SYM?

Table \ref{tab:risk} reports the likelihood of a vulnerability being a threat
if it is contained in one of our datasets. Each row represents a dataset from
which the intersection with the smaller ones has been ruled out: this is to
avoid data overlapping that would falsify the results. The vulnerabilities
which exploits are sold in the market (\EKITS) have 75.73\% chances of
being monitored as actively exploited. This percentage is much lower for \EDB-\EKITS\
and
 \NVD-(\EDB+\EKITS). 
\begin{table}
\begin{center}
\caption{Conditional prob. of vuln. from a dataset being a threat}
\begin{footnotesize}
\noindent Conditional probability that a vulnerability $v$ is listed by
Symantec as threat knowing that it is contained in a dataset, i.e. $P(v\in \SYM
\mid v \in dataset)$.\vspace{5px}
\end{footnotesize}
\label{tab:risk}
\begin{tabular}{|l| c| c| c|}
\hline\hline
            & vuln in \SYM & vuln not in \SYM \\
\hline\hline
\EKITS      & 75.73\% & 24.27\%\\
\EDB-\EKITS &  4.08\% & 95.92\%\\
\NVD-(\EDB+\EKITS) &2.10\%&97.90\% \\
\hline\hline
\end{tabular}
\end{center}
\end{table}

\begin{figure}[t]
\begin{center}
\includegraphics[width=0.40\textwidth]{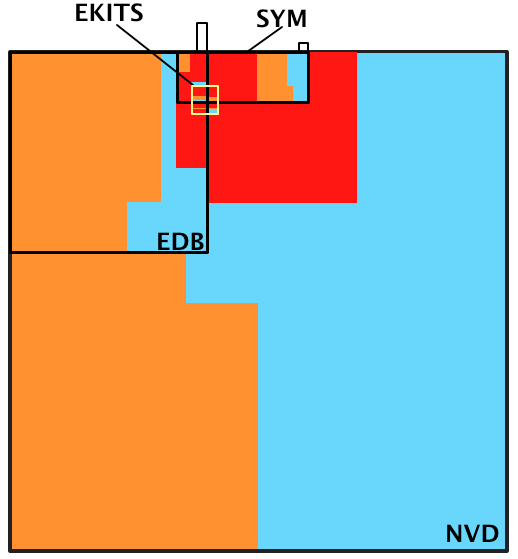}
\end{center}
\begin{footnotesize}
\noindent Dimensions are proportional to data size. In red vulnerabilities with CVSS$\geq$9
score. Medium score vulnerabilities are orange, and cyan represents vulnerability with CVSS lower than 6.
The two small rectangles outside of
\NVD space are vulnerabilities whose CVEs are not present in \NVD.
\end{footnotesize}
\begin{center}
\caption{Relative Map of vulnerabilities per dataset} \label{fig:cvss-map}
\vspace*{-2\baselineskip}
\end{center}
\end{figure}

Figure~\ref{fig:cvss-map} is a Venn diagram of our datasets;
size of the area is proportional to the number of vulnerabilities
and the color is an indication of the CVSS score (a detailed analysis
of the CVSS scores will follow up briefly).

As one can see from the picture many vulnerabilities in the \NVD\ are not
exploited. The \EDB\ is not overly better in terms or representativeness of
actual exploitation in the wild: \EDB\ and \SYM\ share 393 vulnerabilities
only. This means that \EDB\ does not contain about 75\% of the threats measured by
Symantec in the wild. As a minor note, at the collection time
\NVD\ did not reference all vulnerabilities we found:
the \SYM\ and \EDB\ datasets contain respectively 9 and 63 vulnerabilities that
are not present in the \NVD\ dataset. CVSS data on these vulnerabilities is
therefore missing.

A rushing conclusion might be that, if one sees a vulnerability affecting
his/her software in the black market, there is roughly a 75\% chance that it is
exploited in the wild. The same cannot be said about \EDB\ and \NVD, for which
the percentages is less than 5\%. However, a possible counter observation would be that
\EDB\ and \NVD\ include many low CVSS score vulnerabilities and therefore
better results
could be obtained if we eliminate the vulnerabilities with little chances of
being exploited.

\begin{figure}[t]
\begin{center}
\includegraphics[width=0.49\textwidth]{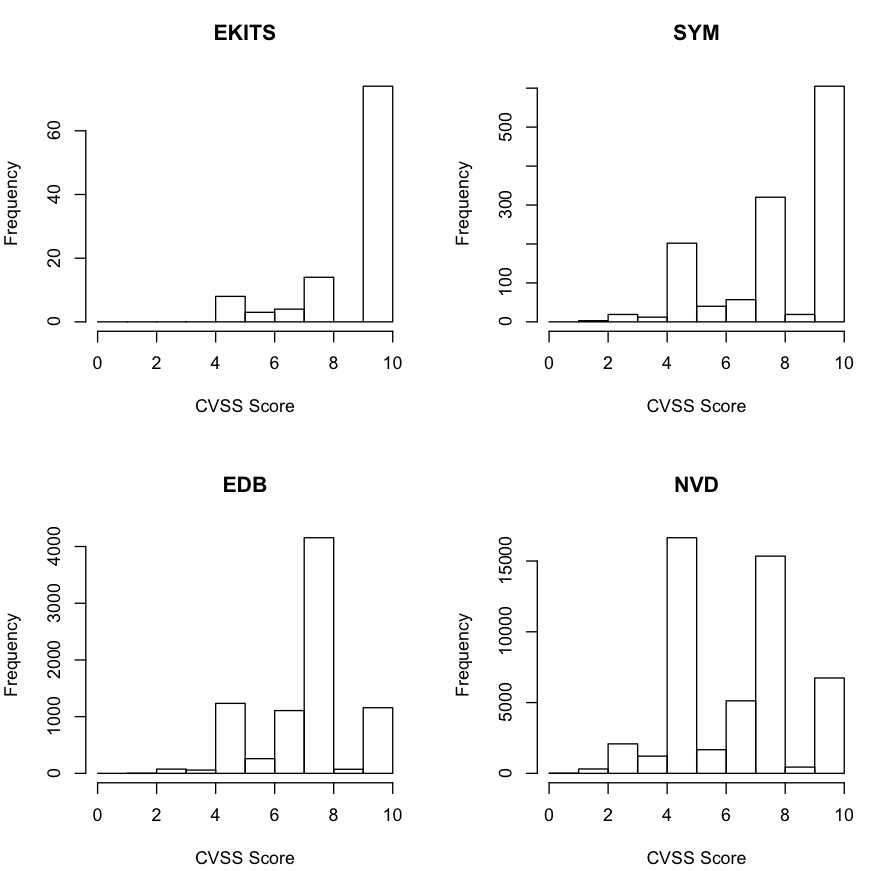}
\caption{Distribution of CVSS scores per dataset.}
\label{fig:hist-scores}
\end{center}
\end{figure}

%
To analyze the CVSS score we report
its histogram distribution in Figure \ref{fig:hist-scores}.
We identify three clusters of vulnerabilities throughout all our
datasets and three corresponding categories of scores:
\begin{enumerate}
\item HIGH: CVSS $\geq$ 9
\item MEDIUM: 6 $\leq$ CVSS $<$ 9
\item LOW: CVSS $<$ 6
\end{enumerate}
In Figure~\ref{fig:cvss-map}, red, orange and cyan areas represent HIGH, MEDIUM
and LOW score vulnerabilities respectively. The \NVD\ reports a large number of HIGH CVSS
vulnerabilities that are not included in \SYM. Similarly, while most of the intersection
between \EDB\ and \SYM\ is covered by HIGH score CVEs, much of the red area for \EDB\
is not included in \SYM.
By looking only at HIGH and MEDIUM
score vulnerabilities in \EDB\ one would deal with about 94\% false positives
(i.e. HIGH and MEDIUM score vulnerabilities not included in \SYM).
False positives decrease to 79\%
if one considers vulnerabilities with HIGH scores only.
Table \ref{tab:cvss-chunks} reports distribution of
HIGH, MEDIUM, LOW scores per each dataset. Looking at \SYM, 52\% of its vulnerabilities
have a CVSS score strictly lower than 9 (665 out of 1277), and 21\% are
strictly lower than 6 (272): 1 out of 5 vulnerabilities exploited in the wild
is ranked as a ``low risk vulnerability'', and 1 out of 2 as ``non-high risk''.
From a first look, HIGH CVSS scores seem to be over-estimating the risk of exploitation
for a large volume of vulnerabilities both in \NVD\ and \EDB, and not being representative of
vulnerabilities  in \SYM.
\begin{table}[t]
\caption{Incidence of CVSS scores per dataset} \label{tab:cvss-chunks}
\centering
\begin{tabular}{|l| r| r| r| r|}
\hline\hline
CVSS Score & EKITS & SYM & EDB & NVD\\
\hline\hline
HIGH & 86 & 612  & 1.209 &7.026\\
MEDIUM & 16 & 393   & 5.324 & 20.858 \\
LOW &12 &272 &1.589   & 21.715 \\
\hline
{\bf tot}& {\bf 114} & {\bf 1.277} & {\bf 8.122} & {\bf 49.599}\\\hline\hline
\end{tabular}
\end{table}


However, this is only an observational analysis from which it is hard to make
statistically sound conclusions: (a) HIGH, MEDIUM or LOW CVSS scores may
only be loosely correlated to inclusion of a vulnerability in \SYM. (b) These results are
strongly influenced by the volume of the datasets: \NVD\ contains almost 50.000
vulnerabilities, while those monitored in the wild are less than 1.300. In \NVD\ and
\EDB\ there might be a lot of ``noisy'' vulnerabilities that are not reported in \SYM\
because of other factors, such as age or software type (i.e. old or rare vulnerabilities
that Symantec may not detect). To
address (a) we measure the goodness of
the CVSS score as a test for exploitation by means of two metrics, namely \emph{sensitivity} and
\emph{specificity} (\S\ref{sec:sensi-speci}). As for (b), we further explore
the CVSS \emph{sub}scores of vulnerabilities to underline technical peculiarities of vulnerabilities in \SYM\
(\S\ref{sec:exploits-characteristics}) and use these as control variables to
sample from \EKITS, \EDB, and \NVD\ vulnerabilities representative of those
reported in \SYM\ (\S\ref{sec:randomized}).

\section{Sensitivity and specificity}\label{sec:sensi-speci}
In the medical domain, the sensitivity of a test is the conditional probability
of the test giving positive results when the illness is present. The
specificity of the test is the conditional probability of the test giving
negative result when there is no illness. In our context, we want to assess to
what degree our current test (the CVSS score) predicts the illness (the
vulnerability being actually exploited in the wild and tracked in \SYM).

Following the preliminary analysis in Section \ref{sec:comparison} we consider
MEDIUM and HIGH CVSS scores as positive tests while LOW scores are negative
tests. In formulae, Sensitivity=$Pr(v.score \geq 6 \mid v \in \SYM)$ while
Specificity= $Pr(v.score < 6 \mid v \notin \SYM)$. Table \ref{tab:spec-sens}
reports the observational specificity and sensitivity for each dataset.
\begin{table}
\begin{center}
\caption{Observational Specificity and Sensitivity of each dataset.}
\begin{footnotesize}
\noindent Sensitivity is the probability of the CVSS score being medium or high
for vulnerabilities actually exploited in the wild. Specificity is the
probability of the CVSS score being low for vulnerability not
exploited in the wild.\vspace{5px}
\end{footnotesize}
\label{tab:spec-sens}
\begin{tabular}{|l| c| c| c|}
\hline\hline
test(v.CVSS) = H v M | \SYM & \EKITS & \EDB & \NVD\\
\hline\hline
Sensitivity & 97.4\%& 94.4\%&78.7\%   \\
Specificity & 32.0\%& 20.3\%&44.4\% \\
\hline\hline
\end{tabular}
\end{center}
\end{table}

For the CVSS score to be a good indicator within a dataset, sensitivity and
specificity should be both high, possibly over 90\%. As shown in Table
\ref{tab:spec-sens}, \EKITS\ is the dataset that performs the best in terms of
sensitivity: out of 100 vulnerability exploited in the wild 97 are
predicted to be dangerous (H or M CVSS score). \EDB\ scores well in terms
of sensitivity too: a proof-of-concept exploit \emph{and} a
HIGH or MEDIUM CVSS score
may be a good test for exploitation. Differently for \NVD, a HIGH or MEDIUM CVSS
score is \emph{not} a good indicator that an exploit will actually show off in
the wild: 21 vulnerabilities out of 100 which are actually dangerous would fail
to get the HIGH or MEDIUM score (79\% sensitivity).
Unfortunately, all databases show poor specificity:
more than 1 out of 2 \emph{not} dangerous vulnerabilities would be wrongly
tagged with a HIGH or MEDIUM score. Loosely speaking, the CVSS test would
generate a medical unnecessary panic among otherwise healthy individuals.



\section{The Impact and Exploitability Subscores}\label{sec:exploits-characteristics}

The general CVSS score takes into consideration two subscores: \emph{Impact}
and \emph{Exploitability}. The former is a measure of the potential damage that
the exploitation of the vulnerability could cause to the victim system; the
latter attempts at measuring the likelihood-to-be-exploited of the
vulnerability \cite{Bozorgi-2010-SIGKDD}. They are calculated on the basis of
further variables that are reported in Table \ref{tab:cvss-metrics}. 
\begin{table}[t]
\centering
\caption{Possible values for the Exploitability and Impact subscores.}
\label{tab:cvss-metrics}
\begin{tabular*}{0.37\textwidth}{@{\extracolsep{\fill}} |c |c| c|}
\multicolumn{3}{c}{Impact subscore}\\\hline
Confidentiality & Integrity & Availability\\
\hline\hline
Undefined & Undefined & Undefined\\
None & None & None\\
Partial & Partial & Partial\\
Complete&Complete&Complete\\\hline
\end{tabular*}
\begin{tabular}{|c |c| c|}
\multicolumn{3}{c}{Exploitability subscore}\\\hline
Access Vector & Access complexity & Authentication\\
\hline
Undefined & Undefined & Undefined\\
Local & High & Multiple\\
Adjacent Net. & Medium & Single\\
Network & Low & None\\\hline
\end{tabular}
\end{table}

\begin{figure}[t]
\begin{center}
\includegraphics[width=0.49\textwidth]{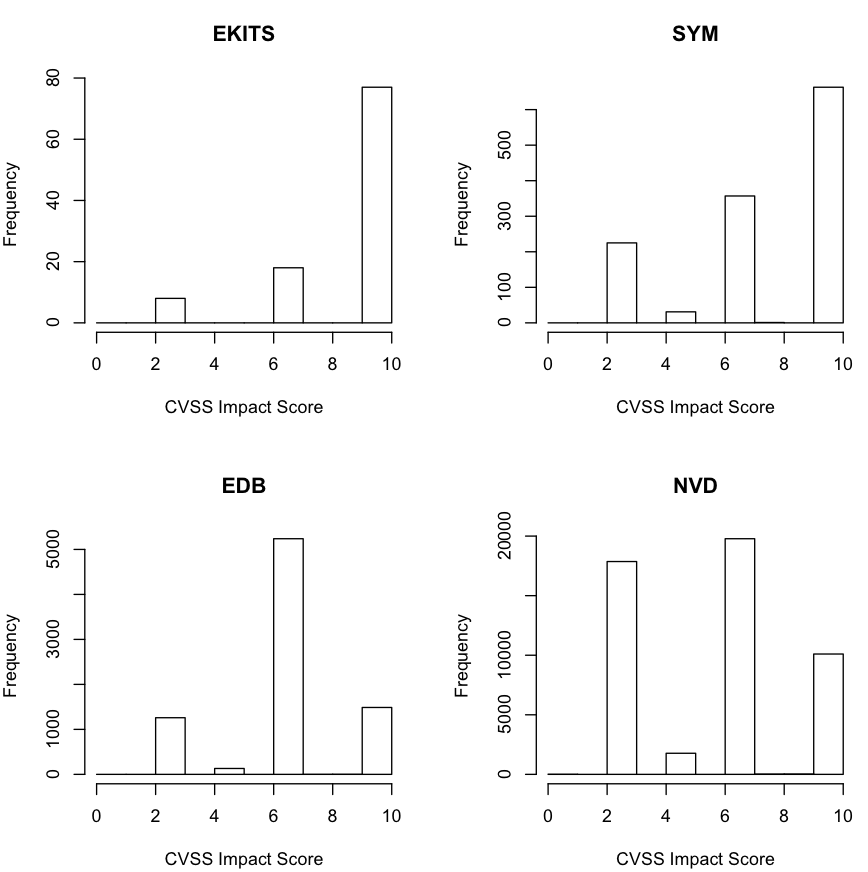}
\caption{Distribution of CVSS Impact subscores per dataset.}
\label{fig:hist-impact}
\end{center}
\end{figure}

The impact metric distribution is plotted in
Figure \ref{fig:hist-impact}.
Somewhat surprisingly, high impact score vulnerabilities are not by default
preferred by attackers: 20\% of vulnerabilities in \SYM\ have a LOW
Impact score. This effect is much reduced for the \EKITS\
dataset: only 8\% of its vulnerabilities score LOW.
As for \EDB\ and \NVD,
the picture changes completely: the greatest majority of vulnerabilities in
\EDB\ (5245, or 65\%) have a medium score, and the remaining 35\% is equally
split between HIGH and LOW Impact vulnerabilities. This might explain the low
specificity for \EDB: many vulnerabilities that just have a
proof-of-concept exploit are of little harm. In \NVD\ 20\%
have HIGH Impact score, 40\%
are scored MEDIUM, and 40\% LOW.



\subsection{Breakdown of the Impact subscore}
Looking in more detail into the Impact metric, Table \ref{tab:availability} shows the
incidence of values of the Confidentiality, Integrity, Availability assessments for
vulnerabilities in the \SYM\ dataset.
Negligible configurations are represented by a handful of vulnerabilities (e.g.
the CCN case is represented by 1 vulnerability).
\begin{table}
\centering \caption{Incidence of values of CIA triad within the \SYM\ dataset.}
\begin{tabular}{|c |c| c| c| c|}
\hline\hline
Confidentiality&Integrity&Availability&\SYM\ &Negligible\\
\hline
C&C&C&51.53\%&\\
C&C&N&0.08\%&\checkmark\\
\hline
C&N&C&0.08\%&\checkmark\\
C&N&N&0.23\%&\checkmark\\
\hline
P&P&P&26.16\%&\\
P&P&N&1.64\%&\checkmark\\
\hline
P&N&P&0.16\%&\checkmark\\
P&N&N&7.67\%&\\
\hline
N&C&C&0.23\%&\checkmark\\
\hline
N&P&C&0.08\%&\\
N&P&P&0.63\%&\\
N&P&N&3.68\%&\\
\hline
N&N&C&1.57\%&\checkmark\\
N&N&P&6.26\%&\\
\hline\hline
\end{tabular}
\label{tab:availability}
\end{table}
Availability almost always assumes the same value as
Integrity, apart from the case where both Integrity and Confidentiality are set
to ``None''. The average variation of the Impact score if Availability was not
to be considered at all is less than 1\%. 

For the sake of readability, we therefore
exclude the Availability assessment from the analysis,
and proceed by looking at the two remaining Impact variables in the four datasets.
The analysis is reported in Table~\ref{tab:conf-int}.
\begin{table}
\centering
\caption{Combinations of Confidentiality and Integrity values per dataset.}
\begin{tabular}{|c |c |c |c |c |c|}
\hline\hline
Confidentiality&Integrity&\SYM&\EKITS&\EDB&\NVD\\
\hline
C&C&51.61\%&74.76\%&18.11\%&20.19\%\\
C&P&0.00\%&0.00\%&0.02\%&0.04\%\\
C&N&0.31\%&0.97\%&0.71\%&0.88\%\\
\hline
P&C&0.00\%&0.00\%&0.01\%&0.01\%\\
P&P&27.80\%&16.50\%&63.52\%&37.84\%\\
P&N&7.83\%&0.97\%&5.61\%&10.62\%\\
\hline
N&C&0.23\%&0.00\%&0.18\%&0.22\%\\
N&P&4.39\%&2.91\%&5.07\%&16.52\%\\
N&N&7.83\%&3.88\%&6.75\%&13.69\%\\
\hline\hline
\end{tabular}
\label{tab:conf-int}
\end{table}
Most vulnerabilities in the \NVD\ dataset score ``partial'' in the two Impact
sub-metrics. This effect is enhanced in the \EDB\ dataset, where almost 70\%
of vulnerabilities score partial in at least one of either Confidentiality
or Integrity. The scenario changes completely when looking at the
\SYM\ and \EKITS\ datasets: most vulnerabilities (~50\%, ~75\%) score
``Complete''. Black market exploits and exploits in the wild seem to affect
vulnerabilities with different impact types than the general population
of vulnerabilities and proof-of-concept exploits.

%


\subsection{Breakdown of the Exploitability subscore}
\begin{figure}[t]
\begin{center}
\includegraphics[width=0.49\textwidth]{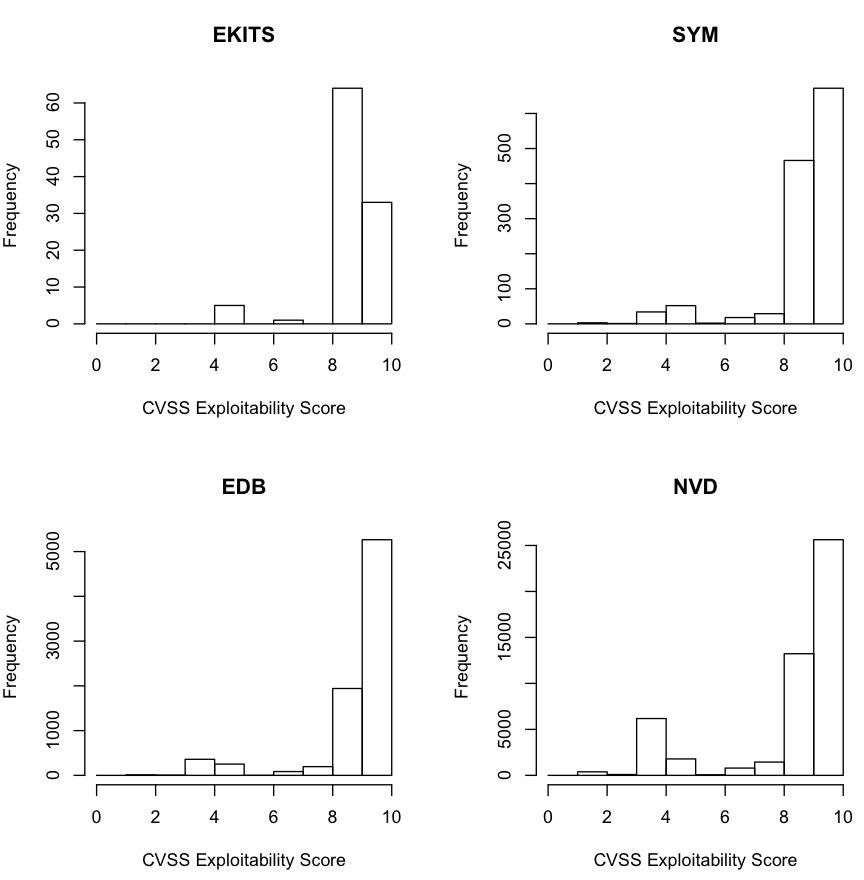}
\caption{Distribution of CVSS Exploitability subscores.}
\label{fig:hist-expl}
\end{center}
\end{figure}

Figure \ref{fig:hist-expl} shows the distribution of the Exploitability
subscore per each dataset.
Here the distinction between HIGH and MEDIUM Exploitability scores seem to be
not very significant: numbers are qualitatively identical among all datasets.
Almost all vulnerabilities, independently of the dataset, score between 8 and 10.
Bozorgi et al.'s \cite{Bozorgi-2010-SIGKDD} did not find any correlation between Exploitability subscore and existence of a proof-of-concept
exploit; we confirm their results and extend this conclusion to 
exploits in the wild and exploits in the black markets as well. 

The Exploitability subscore resembles more a constant than a variable.
This means that it has little to no influence on the variance of the final CVSS
score, which may in turn affect the suitability of the CVSS as a \emph{risk} metric.

Table \ref{tab:cvss-expl} reports the total distribution of the exploitability
variables.
\begin{table}[t]\scriptsize
\centering \caption{Exploitability Subfactors for each dataset.}
\label{tab:cvss-expl}
\begin{tabular}{|l| l| l| c| c| c| c|}
\hline\hline
& metric& value & \SYM & \EKITS & \EDB & \NVD\\
\hline\hline
\multirow{9}{*}{\begin{sideways}Exploitability\end{sideways}}
	& \multirow{3}{*}{Acc. Vec.}& local & 2.98\% &0\%& 4.57\%&13.18\%\\
	&	&adj.&0.23\%&0\%&0.12\%&0.35\%\\
	&	&net&96.79\%&100\%&95.31\%&87.31\%\\\cline{2-7}
	& \multirow{3}{*}{Acc. Com.} & high &4.23\%&4.85\%&3.37\%&4.54\%\\
	&	&medium&38.35\%&63.11\%&25.49\%&30.42\%\\
	&	&low&57.24\%&32.04\%&71.14\%&65.68\%\\\cline{2-7}
	& \multirow{3}{*}{Auth.}&multiple&0\%&0\%&0.02\%&0.05\%\\
	&	&single&3.92\%&0.97\%&3.71\%&5.35\%\\
	&	&none&96.08\%&99.03\%&96.27\%&95.45\%\\\hline\hline
\end{tabular}
\end{table}
The greatest share of actual risk comes from vulnerabilities that can be
remotely exploited; just 3\% of vulnerabilities are only locally
exploitable. Moreover, the great majority of discovered vulnerabilities is
network-based (87.31\%). Authentication is another essentially boolean
variable: most exploited vulnerabilities do not require any authentication.

\section{Exploitation Trade-offs} \label{sec:exploitation-tradeoff}

Among all subscores, access complexity present some interesting results: the
percentage of ``very difficult'' vulnerabilities is equal (and very low) among
all datasets but the percentage of ``medium-complexity'' vulnerabilities in the
\SYM\ and \EKITS\ datasets is much higher than in \EDB. 
Medium-complexity
vulnerabilities in the \EKITS\ and \SYM\ datasets are respectively 63.11\% and
38.35\% of the totals. As a comparison, only 25.49\% of vulnerabilities in the
\EDB\ dataset have medium-complexity. Exploits in \EDB\ seem to mostly capture easy
vulnerabilities (71.14\%).


To explain the higher average Complexity for vulnerabilities exploited in the
wild we hypothesized a trade-off for the attacker: he/she is willing to put
extra-effort in the exploitation only if it is worth it (i.e. the vulnerability has HIGH Impact).
Table \ref{tab:cvss-trade-off} reports the results of the analysis.
\begin{table}[t]\scriptsize
\centering
\caption{Relationship between Access Complexity, Impact and actual exploitation}
\begin{tabular}{|l l |l| c| c| c| c|}
\hline\hline
&  & Impact &\SYM & \EKITS & \EDB & \NVD\\
\hline\hline
\multirow{9}{*}{\begin{sideways}Access Complexity\end{sideways}}
	& \multicolumn{1}{|c|}{\multirow{3}{*}{High}} & HIGH &1.33\%&2.91\%&0.58\%&0.92\%\\
		&  \multicolumn{1}{|c|}{} &MEDIUM&1.88\%&1.94\%&2.34\%&1.89\%\\
		& \multicolumn{1}{|c|}{} & LOW & 1.02\%&0.00\%&0.46\%&1.89\%\\\cline{2-7}
	& \multicolumn{1}{|c|}{\multirow{3}{*}{Medium}} & HIGH &32.50\%&55.34\%&8.84\%&7.65\%\\
		& \multicolumn{1}{|c|}{} & MEDIUM & 3.60\%&4.85\%&11.35\%&7.69\%\\
		& \multicolumn{1}{|c|}{} & LOW&2.43\%&2.91\%&5.29\%&14.83\%\\\cline{2-7}
	& \multicolumn{1}{|c|}{\multirow{3}{*}{Low}} & HIGH & 18.09\%&16.50\%&8.89\%&11.80\%\\
		&\multicolumn{1}{|c|}{}  &MEDIUM & 22.55\%&10.68\%&50.89\%&30.43\%\\
		&\multicolumn{1}{|c|}{}  &LOW&16.60\%&4.85\%&11.36\%&22.90\%\\ \hline\hline
\end{tabular}
\label{tab:cvss-trade-off}
\end{table}
The trade-off is particularly evident in the medium-complexity range of
vulnerabilities: if an attacker is going to exploit a medium complexity
vulnerability, most likely this will be a HIGH impact one (32.50\%). This trend
is even more evident in the \EKITS\ dataset, in which this percentage increases
to 55.34\%. This supports the hypothesis that the extra effort required to
write an exploit for a more complex vulnerability is to be weighted with a
corresponding ``return on investment''. Upon LOW Complexity vulnerabilities, on
the other hand, there is no clear difference between HIGH, MEDIUM and LOW
impacts: as long as exploitation is easy, the attacker may be willing of
exploiting it regardless of the Impact score.

\section{Randomized Case-Controlled study}
\label{sec:randomized}

In order to obtain stronger statistical results on the suitability of the CVSS score as a risk metric for
vulnerabilities, we 
generate a
case-controlled study where the cases are the vulnerabilities in \SYM\
(loosely corresponding to cases of lung cancer), and \NVD, \EDB, and
\EKITS\ correspond to patients from various sources and medical conditions. We
are looking for a control variable (like smoking) that could overwhelmingly
explain exploitation (cancer).

\emph{\bf Confounding variables.}
Vulnerabilities in \EKITS, \EDB\ and \NVD\ may be generally different
from those in \SYM\ with respect to a number of factors. For example,
vulnerabilities in \EDB\ may contain software that is of no \emph{commercial
interest} for Symantec, and therefore for which an entry in \SYM\ cannot exist.
Another possibility is that the chances for a vulnerability being reported in \SYM\ 
may vary in time (for example, because less attacks existed in that year, or
because Symantec may not always perform equally well
in detecting attacks). 

We identify as possible control variables for inclusion in \SYM\
\emph{confidentiality,
integrity, availability, software and year}. We then
generate three random samples of vulnerabilities
from the \EKITS, \NVD\ and \EDB\ datasets by selecting vulnerabilities
which characteristics (i.e. our confounding variables) 
appear in \SYM\ as well.
We report here the results with the highest statistical significance, 
those resulting from the samples controlled by Impact.
Additionally, after feedback from the BlackHat community, in Appendix~\ref{appendix:controls}
we report how results change by considering the other confounding variables, alongside
with confidence intervals and statistical significance (p-value).

\emph{\bf Sampling.} 
Random sampling
is necessary because our data
may not reflect the complete population of (a) disclosed vulnerabilities; (b) existing
proof-of-concept exploits and exploits traded in the black markets; 
(c) vulnerabilities actually exploited in the wild.

However, randomness may introduce (random) noise in the analysis.
To measure this (and draw more precise conclusions) we perform the sampling 
with \emph{bootstrapping}:
we bootstrap our experiment by sampling, with repetition, 
400 times for each control and each dataset
\footnote{The sampling was performed with the
statistical tool R-CRAN~\cite{R-CRAN}.}.

\emph{\bf Results.}
Table~\ref{tab:fisher-sensi} reports the median, for each sample, of the results. We
consider as a (tentative) explanatory variable the value of the CVSS and as
response variable the presence of the vulnerability in \SYM. 
\begin{table}
\begin{center}
\footnotesize \caption{Case-controlled Conditional Probability}
\begin{footnotesize}
\noindent Case-controlled distribution among dataset of CVSS scores
(explanatory variable) vs actual exploit in the wild as reported by \SYM\
(response variable).\vspace{5px}
\end{footnotesize}
\label{tab:fisher-sensi}
\begin{tabular*}{0.49\textwidth}{@{\extracolsep{\fill}}|l| c| c| c|}
\multicolumn{4}{c}{\SEKITS} \\
\hline
              & v $\in$ \SYM  & v $\notin$ \SYM &p-value \\
\hline
 CVSS H or M    & 76 (81.72\%) & 17 (18.28\%) &\multirow{2}{*}{$1.45\exp{-4}$}\\
 CVSS L            & 2 (20\%) & 8 (80\%) &\\
\hline
\multicolumn{4}{c}{\SEDB\rule{0pt}{3ex}} \\
\hline
              & v $\in$ \SYM  & v $\notin$ \SYM &p-value \\
\hline
CVSS H or M  & 104(11.84\%) & 774 (88.15\%)  &\multirow{2}{*}{$2.46\exp{-8}$}\\
CVSS L         &   3 (1.28\%) & 232 (98.72\%)  &\\
\hline
\multicolumn{4}{c}{\SNVD\rule{0pt}{3ex}} \\
\hline
              & v $\in$ \SYM  & v $\notin$ \SYM &p-value \\
\hline
CVSS H or M     & 48 (5\%) &  912 (95\%)&\multirow{2}{*}{$6.10\exp{-3}$}\\
CVSS L    & 4  (1.38\%) & 285 (98.62\%)& \\
\hline
\end{tabular*}
\end{center}
\end{table}

To check for statistical significance of the data 
we run a Fisher's exact test (because data is not normal) for each outome.
The p-values are reported in Table \ref{tab:fisher-sensi}. We recall that the $p$
value does not measure the strength of an effect or an association (it is up to
us to see it in the data), but only the certainty that the effect that we see
in the data is not due to chance. A $p$ value less than 0.05 is considered
statistically significant because there is less than 5\% chances that the data
could show the distribution by chance.


\subsection{Sensitivity and specificity for case-controlled study}

Table~\ref{tab:spec-sens:controlled} reports the median sensitivity and specificity
of the CVSS score for the controlled samples. 95\% confidence intervals
are reported within square brackets. The 0\% 
variation in sensitivity and specificity of the \EKITS\ dataset is to be attributed to the small relative
size of \EKITS\ vs \SYM: in the sampling, all vulnerabilities from \EKITS\ are chosen. Sensitivity is quite high among all
the datasets. 
Yet, the CVSS score has a  poor specificity for all datasets. Sampling
\SYM-like characteristics does not help in scoring vulnerabilities as
``non-dangerous'' ones. In particular, if we consider a specificity of 25\%, only 1 out of 4 
non-attacked
vulnerabilities are marked as LOW score. 
The remaining three are instead marked as MEDIUM or HIGH.
\begin{table}
\begin{center}
\scriptsize
\caption{Case-controlled Specificity and Sensitivity.}
\begin{footnotesize}
\noindent Case-controlled sensitivity and
specificity of the CVSS score being medium or high and the vulnerability being
actually exploited in the wild (i.e. in \SYM). Data has been random sampled
from \EDB\ and \NVD\ according \SYM's distribution of values for CVSS
subscores. 95\% confidentiality interval is here reported in square brackets [].\vspace{5px}
\end{footnotesize}
\label{tab:spec-sens:controlled}
\begin{tabular}{|l| c| c| c|}
\hline
 & \EKITS & \SEDB & \SNVD\\
\hline
Sens. &97.44\% [$\pm$ 0\%] &97.17\% [94.33\%-100\%]&91.88\% [84.20\%-98.11\%]\\
Spec. &32\% [$\pm$ 0\%]&23.09\% [21.94\%-24.33\%]& 23.82\% [22.52\%-25.02\%]\\
\hline
\end{tabular}
\end{center}
\end{table}

From table \ref{tab:spec-sens:controlled} we also conclude that the CVSS score
performs differently according to the reference dataset used.

\subsection{Rule-based policies for risk mitigation with CVSS}

From our analysis, using the CVSS score as a test for exploitation does not provide enough
precision for effective patching policies.
However, this only outlines the \emph{a-priori} performance of the CVSS test in identifying exploits,
and says little about the effectiveness of it as a ''metric for patching''.
We here address the question: \emph{What is the marginal diminishment
of risk provided by CVSS when coupled with different datasets?}


To understand wether a \emph{HIGH CVSS $\rightarrow$ HIGH risk} policy is meaningful,
we adopt an approach similar to that used by Evans in  \cite{Evans-1986-ACC}
to estimate the effectiveness
of seat belts in preventing fatalities. In his case,
the ``effectiveness'' was given by the difference in the probability of having a fatal
car crash when wearing a seatbelt and when not. More formally, Pr(Death x Seat belt)
- Pr(Death x not Seat belt).
In our case, the effect we are interested in is the ability of
the CVSS score (combined with the datasets) to predict the actual exploit in the
wild (i.e. present in \SYM). Table~\ref{tab:sample-diff-ratio} shows
the difference of the probabilities.
\begin{table}
\centering \caption{Relative Risk for CVSS score}
\vspace*{-\baselineskip}
\begin{footnotesize} \noindent Increase in relative risk for a vulnerability
to be exploited depending on CVSS score, database and control variables used
for the study i.e.$P(\SYM|H+M)-P(\SYM|L)$.\vspace{5px}

\end{footnotesize}
\begin{tabular}{|l |c| c|}
\hline
\hline
Database & Results & 95\% Conf. Interval \\
\hline
\SEKITS & +61.72\% &$\pm$ 0\%\\
\SEDB   & +10.51\% & 8.39\%-12.64\%\\
\NVD    &  +3.54\% &1.60\%-5.28\%\\
\hline\hline
\end{tabular}
\label{tab:sample-diff-ratio}
\end{table}

Each row in the table tells us the
chances that a vulnerability with a MEDIUM or HIGH CVSS score is actually
exploited in the while vs one with LOW scores.

For \SEKITS\ we see that fixing HIGH/MED vulnerabilities from the black markets
carries a decrease in risk by +61\% with respect to the immediate lower level
of risk (i.e. the remaining vulnerabilities after the patching). 
For \SEDB, the evidence is
less strong. We only have +10\% more chances of exploitation
for the high level of risk than for the low level. 
Finally, \SNVD\ has even weaker association for the same
reasons: we only have +3\% increase in chances of exploitation for HIGH/MED
vulnerabilities.
We conclude that rule-based policies based solely on the CVSS score 
 may result inadequate for
deployment of effective security countermeasures.
Coupling the CVSS test with other evidence carries more informative results than
simply using the score as-is.

\section{Discussion and implications}\label{sec:discussion}

Vulnerability assessment and patching has traditionally been a matter of great
discussion within the community
\cite{Clark-2010-ACSAC,Schryen:2011:OSS:1941487.1941516,Shahzad-2012-ICSE}.
Here we summarize the main implications from our study.

\textbf{Implication \#1.} Vulnerabilities exploited in the wild show specific
patterns in the CVSS subscores; these observations can help to improve the
sensitivity and specificity of the CVSS score. Some conclusions are more
absolute (exceptions counted on one's fingers), while others are only
statistically significant (hence the adverb ``usually''), with a $p$-value lower
than $<2.2E-16$ for Fisher's exact test.
\begin{enumerate}
\item \emph{Actually exploited vulnerabilities are remotely exploitable and
    do not require multiple authentication}. Despite \SYM\ containing local
    threats, only 3\% of vulnerabilities are assessed as ``only locally
    exploitable''. Vulnerabilities exploitable from an adjacent network are
    even less interesting. 4\% of vulnerabilities require a single instance
    of authentication; none of them require multiple authentication.
\item \emph{Availability impact is irrelevant}. The impact of more than
    96\% of vulnerabilities in \SYM\  can still be accurately assessed
    without taking into consideration the value of Availability.
\item \emph{Confidentiality and Integrity losses usually go hand-in-hand}.
    The overwhelming majority of vulnerabilities in \SYM\ have complete or
    partial losses for both Confidentiality and Integrity: other
    combinations are less likely to be exploited. 
\item \emph{``Exploits'' in \EDB\ are usually for easy vulnerabilities}.
    Proof-of-concept exploits released in the \EDB\ are for vulnerabilities
    easier to exploit than those actually exploited by attackers.
\item \emph{Medium-complexity vulnerabilities are usually interesting only
    if they come along with a high impact}. Non-trivial to exploit
    vulnerabilities
seem to be of interest for the attacker only if they come with a higher
final impact on the vulnerable system. In contrast, Low-complexity vulnerabilities
    are exploited uniformly among all impact scores.
\end{enumerate}

\textbf{Implication \#2.} Rule-based policies based on CVSS score, like the US
Government NIST SCAP protocol \cite{Scarfone-2010-SCAP}, may not make for an
effective strategy: only a negligible number of low-risk vulnerabilities are
ruled out, even after controlling for ``significant'' vulnerabilities. Security
policies may require a major adjustment to meet these observations. In
particular, while the CVSS score underlines interesting characteristics of
exploited
    vulnerabilities,
it may be not expressive enough to reliably represent exploitation. Other
    factors such as software popularity, presence of the exploit in the
    market and existence of easier vulnerabilities for that software are
    all ``contextual factors'' that might be worth exploring in future work.

\textbf{Implication \#3.} The black market can be a good source to assess which
vulnerabilities represent high risk. Exploits for vulnerabilities traded in the
black market significantly overlap with those recorded in the wild, which may indicate
that the presence of an exploit in the black markets
can be a good indicator of the associated vulnerability risk.

\section{Threats to validity}\label{sec:threats}

We identify a number of threats to validity.
\cite{Perry-ICSE-2000}.

{\bf Construct validity} 
A number of issues we encountered while collecting \SYM\ and \EKITS\ may affect
the soundness of the data collection.
Because of the unstructured dataset of the original SYM dataset, to build \SYM\
we needed to take some preliminary steps.
We couldn't be
sure about whether the collected CVEs were relevant to the threat. To address
this issue, we proceeded in two steps. First, we manually analyzed a random selection
of about 50 entries to check for the relevance of the CVE entries in the
``description'' and ``additional references'' sections of each entry.
To double-check our evaluation, we questioned Symantec in an
informal communication: our contact confirmed that the CVEs are indeed
relevant.
Another issue is what data from Symantec's attack-signature and
threat-explorer datasets to use. Attack and infection dynamics are not always
straightforward, and network and host-based threats often overlap. However, in
this case, we are interested in a general evaluation of risk. Moreover, Exploit
Kits enforce a drive-by download attack mechanism, therefore they are related
to both the network and local threat scenario. We therefore can safely rely on
both the datasets for our analyses.

Due to the shady nature of the
tools, the list of exploited CVEs in \EKITS\ may be incomplete and/or incorrect. We don't
know any straightforward way to address this issue; to mitigate the problem, we
crossed-referenced entries with knowledge from the security research community
and from our direct observation of the black markets.

{\bf External validity} is concerned with the applicability of our results to
real-world scenarios. Symantec is a world-wide company and a leader
in the security industry. We are therefore confident is considering their data
representative sample of real-world scenarios. Yet, our conclusion cannot be
generalized to the risk due to targeted attacks. Targeted attacks in the wild
of a specific platform or system are less likely to generate an entry into a
general anti-virus product, and therefore less likely to be represented in the
\SYM\ database.

\section{Related works} \label{sec:related}

Many studies before ours analyzed and modeled trends in vulnerabilities.
Among all, Frei et al. \cite{Frei-2006-LSAD} were maybe the first to link
the idea of life-cycle of a vulnerability to the patching process. Their dataset
was a composition of \NVD, OSVDB and `FVDB' (Frei's Vulnerability DataBase, obtained
from the examination of security advisories for patches).
The life-cycle of a vulnerability
includes discovery time, exploitation time and patching time. They showed that,
according to their data, exploits are often quicker to arrive than patches are. They were
the first to look, in particular, at the difference in time between time of first ``exploit''
and time of disclosure of the vulnerability. This work have recently been extended by
Shahzad et al. \cite{Shahzad-2012-ICSE}, which presented a
comprehensive vulnerability study on NVD and OSVDB datasets (+ Frei's) that included
vendors and software in the analysis. Many interesting
trends on vulnerability patching and exploitation are presented, and support Frei's conclusion.
However, they basically looked at the same data: looking at \EDB\ or OSVDB
may say little about actual threats
and exploitation of vulnerabilities. The difference with our paper, here,
is that we look at a \emph{sample of actual attack data} (\SYM) and underline differences
in vulnerability characteristics with other datasets. For a thorough description of our datasets
and a preliminary discussion on the data, see \cite{allodi-2012-BADGERS}.
An analysis of the distribution of CVSS scores and subscores
has been presented by Scarfone et al. in \cite{Scarfone:2009} and Gallon \cite{5720656}.
However, while including CVSS subscore analysis,
their results are limited to data from \NVD\ and do not provide any insight on
vulnerability exploitation.
In this sense, Bozorgi et al. \cite{Bozorgi-2010-SIGKDD} were probably the first
in looking at CVSS subscores against exploitation. They showed that the ``exploitability''
metric, usually interpreted as ``likelihood to exploit'' did not match with data from \EDB:
their results were the first to show
that the interpretation of CVSS metrics might not be entirely straightforward. We extended
their first observation with a in-depth analysis of subscores and of actual exploitation data.

On a slightly different line of research are studies concerned with the discovery of
vulnerabilities. In \cite{Clark-2010-ACSAC} Clark et. al. underlined the presence of a `honeymoon effect'
in the discovery of the first vulnerability for a software, that
is related with the ``familiarity'' of the product. In other words,
the more popular the software the smaller the gap between software release and first
vulnerability disclosure. 

Other studies focused on the modeling of the vulnerability discovery processes.
Foundational in this sense are the works of Alhazmi et al. \cite{Alhazmi-2008-TR} and
Ozment's \cite{Ozment:2007:IVD:1314257.1314261}. The former fits 6 vulnerability
models to vulnerability data of four major operative systems, and shows that Alhazmi's `S shaped'
model is the one that performs the better. However, as previously underlined by Ozment
\cite{Ozment:2007:IVD:1314257.1314261}, vulnerability models often rely on
unsound assumptions such as the independence of vulnerability discoveries.
Current vulnerability discovery models are indeed
not general enough to represent trends for all software
\cite{Massacci-2012-ASIACCS}. Moreover, vulnerability disclosure and
discovery are complex processes
\cite{Ozment-2005-WEIS}, and can be influenced by \{black/white\}-hat
community activities \cite{Clark-2010-ACSAC, Frei-2006-LSAD} and economics
\cite{Miller-2007-WEIS}.




Our analysis of the vulnerabilities marketed in exploit-kits is also
interesting because it confirms that the market for exploits is significantly
different than the IRC markets for credit cards and other stolen goods.
Indeed, dismantling some previous analysis \cite{Franklin-2007-CCS}, Herley et
al. \cite{Herley-2010-EISP} shown that IRC markets feature all the
characteristics of a typical ``market for lemons'' \cite{Akerlof1970}: the
vendor has no drawbacks in scamming the buyer because of the complete absence
of a unique-ID and of a reputation system. Moreover, the buyer cannot in any
way assess the quality of the good (e.g.
the amount of credit available) beforehand.

In contrast, Savage et al. \cite{Savage-2011-ICM} analyzed the private messages
exchanged in 6 underground forums. Most interestingly, their analysis shows
that these markets feature the characteristics typical of a regular market:
sellers do re-use the same ID, the transactions are moderated, and reputation
systems are in place and seem to work properly. These observations coincide
with our direct exploration of the black markets. The
results reported in this paper show that by buying exploit kits one buys
something that might actually work: the exploits in exploit kits are actually
seen in the wild.

\section{Conclusion}\label{sec:conclusions}
In this paper we presented our four datasets of vulnerabilities (\NVD),
proof-of-concept exploits (\EDB), exploits traded in the black market (\EKITS),
and exploits recorded in the wild(\SYM). We showed that, in general, the CVSS
score and its submetrics capture some interesting characteristics of the
vulnerabilities whose exploits are recorded in the wild but it is not
expressive enough to be used as a reliable test for exploitation (with both
high sensitivity and high specificity). 

Alas, the bottom-line answer to the question set out in the title of this paper
is not entirely satisfactory. If we compare the results of
Table\ref{tab:sample-diff-ratio} with the risk reduction obtained by wearing
safety belts in cars (43\% according to Evans \cite{Evans-1986-ACC}) \emph{You should
surely worry in few cases:}
\begin{itemize}
\item your vulnerability is listed by an exploit kit in the black market
    and have a medium-high CVSS score (61\% risk reduction);
\item your vulnerability has a proof of concept exploit (eg in \EDB),
    requires no authentication, can be exploited over the network and have
    medium complexity but high-impact (with a medium-high CVSS score) (10\%
    risk reduction).
\end{itemize}

Unfortunately, nor CVSS subscores, nor the existence of exploits, nor the
trading on the black market offer a statistically sound test for ruling out the
98\% of the cases for which users at large shouldn't worry.

Overall we can conclude that \emph{CVSS-alone rule-based policies are not
economically effective}: the compliance with current regulations only yields
low gains in terms of actual security (i.e. 3-4\% risk reduction).


A claim can be made for the databases subject of this study: \emph{using \NVD,
\EDB\ (or consequently OVSDB) to assess software exploits in the wild is the
wrong thing to do.} Without additional attention, those databases can only be
used to assess the upper hand in the race between software vendors and security
researchers.

%
\section*{Acknowledgements}
This work was partly supported by the
EU-SEC-CP-SECONOMICS
and MIUR-PRIN-TENACE Projects. We
would like to thank Tudor Dimitras and Viet H. Nguyen for the many useful
discussions. Viet actually wrote the second script used for cross-checking the
\SYM\ dataset. No statement in this paper should be interpreted as endorsed by
Symantec.

\bibliographystyle{abbrv}
\bibliography{../../../../Biblio/short-names,../../../../Biblio/bibfile}

\newpage
\appendix
\label{appendix:controls}
In this Appendix we report the full set of results from the bootstrapped random sampling described
in Section \ref{sec:randomized}.

The resulting Table \ref{tab:appendix} reports the data for four ``primary'' controls, namely 
\emph{censorYr, censorSw, checkYr, checkSw} $\times$ five ``supplementary'' controls, namely 
\emph{nvd, edb, ekits, edbNOekits, nvdNOekitsNOedb}. This gives us $2^4 \times 5 =80$ 
experiments to report. A fifth ``primary'' control is always active: the CIA CVSS 
assessment of the vulnerability. The remaining four ''primary'' control variables are as follows:

\begin{itemize}
\item censorYr: only CVEs disclosed in the years 2009,2010,2011,2012 are considered in the study.
\item censorSw: only CVEs affecting software also included in \SYM\ are considered. This assures
that our samples include only software monitored by Symantec.
\item checkYr: for each CVE in \SYM\ another is picked for inclusion in the sample that has the \emph{exact same year} as the one in \SYM. This is different from censorYr as in that case the only
constraint is that the chosen vulnerability lays in the 2009-2012 time range.
\item checkSw: each CVE picked for the sample has exactly one counterpart, software-wise, 
in \SYM.
\end{itemize}

The five ``supplementary'' controls are as follows:
\begin{itemize}
\item nvd: the vulnerability has been disclosed (i.e. we consider every vulnerability).
\item edb: the vulnerability has a proof-of-concept exploit in \EDB.
\item ekits: an exploit is reported to be traded in the black markets for that vulnerability.
\item edbNOekits: the vulnerability has a proof-of-concept exploit \emph{and} does not have
an exploit traded in the black markets
\item nvdNOekitsNOedb: the vulnerability is only disclosed, but has no proof-of-concept or exploit
traded in the black markets.
\end{itemize}

The last two ``supplementary'' controls are in particular useful to measure the decrease in risk
of a particular patching policy \emph{after} having applied another one. For example, if one knows
that patching all HIGH score vulnerabilities in \EKITS\ yelds a 60\% risk reduction, by patching the 
remaining vulnerabilities with a proof-of-concept exploit in \EDB\ his/her risk would decrease by only 
an \emph{additional} 10\%. With \NVD, finally, the extra step in risk reduction is of 3\%.

Sensitivity, specificity and risk reduction are calculated on the median of the results obtained from
the bootstrapping. Additionally, we report the 95\% confidence intervals (C.I.) for these measures.
The last column of the table reports the median p-value of the results: the reported p-value
is the value under which 50\% of our samples are. This gives an indication of the statistical
significance for the particular combination of ``primary'' and ``supplementary'' control variables.
We remember the reader that a value lower than 0.05 
is considered to be the limit for statistical significance.

First be noted, because of the limited size of \EKITS, it may happen that a particular
set of control variables leads to both variables in a row in the latin square (see 
Table~\ref{tab:fisher-sensi}) to be zero. In these cases the risk reduction cannot be computed 
because the probability for that row cannot be calculated.
Interestingly, the ``checkSw'' control seems to introduce a lot of statistical noise in the data: when
this flag is activated, the statistical significance of the samples decreases sensibly. This may be 
because the ``software'' attribute may not be a good control variable \emph{as-is}. For example,
consider a vulnerability for \emph{Webkit}, an HTML engine used in many browsers. This vulnerability may affect not only Webkit, but also Safari, Chrome or Opera that
use it as a rendering engine. A different categorisation for software 
may therefore make sense to enforce. For example, the Webkit vulnerability may be included in
a macro-category of \emph{Browser} vulnerabilities from which to sample a corresponding
control vulnerability.

While the considerations for sensitivity and specificity of the CVSS score remain substantially
the same regardless of the control variables, the results for the risk reduction column 
can vary. In particular, by looking at the risk reduction confidence intervals, the risk reduction in 
some cases, for \NVD, may even be \emph{negative}. In these cases it means that patching
HIGH or MEDIUM CVSS score vulnerabilities would make no sense at all, as it would be more
likely to be attacked by a LOW score vulnerability. 
\EDB\ performs particularly well when the controls ``checkYr'' and 
``checkSw'' are activated, achieving a maximum of 46\% risk reduction.
The results in risk reduction for
the CVSS test plus \EKITS\ remain very positive, ranging from +55\% to +80\% with high or good
statistical significance. When considering only ``exclusive'' vulnerabilities (i.e. edbNOekits, nvdNOedbNOekits) figures become slightly lower. This is because, in this case, we are
excluding from the sample the vulnerabilities from \EKITS\ and \EDB+\EKITS\ that have
already been considered for patching (as patching twice the same vulnerability would not
provide any additional reduction in risk).

\begin{table*}
\caption{Full table of results for the bootstrapped case controlled study. The medianPvalue
column reports (***) for p$<$0.001, (**) for p$<$0.01, (*) for p$<$0.05, and nothing for any p$\geq$0.05.}
\label{tab:appendix}
\centering
\tiny
\begin{tabular}{|c |c |c |c |c |c |c |c |c |c |c |c |c |c |c |c |}
\hline
db&censorYr&censorSw&checkYr&checkSw&HvM+Sym&L+Sym&HvM+NotSym&L+NotSym&Sens.&95\% C.I.&Spec.&95\% C.I.&Risk red.&95\% C.I.&p-value\\ \hline\hline
nvd&&&&&48&4&912&285&91.89\%&84.21-98.11&23.82\%&22.53-25.02&3.54\%&1.6-5.28&**\\ 
edb&&&&&104&3&774&232&97.17\%&94.34-100&23.09\%&21.94-24.33&10.51\%&8.39-12.64&***\\ 
ekits&&&&&76&2&17&8&97.44\%&97.44-97.44&32.00\%&32-32&61.72\%&61.72-61.72&***\\ 
edbNOekits&&&&&88&3&788&231&96.77\%&93.46-100&22.73\%&21.58-23.99&8.79\%&6.54-10.74&***\\ 
nvdNOekitsNOedb&&&&&35&4&905&301.5&88.89\%&78.12-97.62&25.03\%&23.77-26.22&2.27\%&0.36-3.96&\\ 
\hline nvd&$\checkmark$&&&&43&2&891&265.5&94.87\%&86.48-100&22.98\%&21.94-23.8&3.59\%&1.71-5.11&**\\ 
edb&$\checkmark$&&&&88&1&648&204&98.98\%&97.7-100&23.95\%&22.67-25.15&11.77\%&10.13-13.31&***\\ 
ekits&$\checkmark$&&&&40&0&10&3&100.00\%&100-100&23.08\%&23.08-23.08&80.00\%&80-80&*\\ 
edbNOekits&$\checkmark$&&&&71&0&657&204&100.00\%&97.22-100&23.71\%&22.58-24.83&9.41\%&8.15-10.85&***\\ 
nvdNOekitsNOedb&$\checkmark$&&&&26&2&880&277&91.37\%&78.94-100&23.99\%&22.79-25.17&1.97\%&0.3-3.39&\\ 
\hline nvd&&$\checkmark$&&&125&17&762&299&87.99\%&82.58-92.48&28.17\%&26.73-29.81&8.53\%&5.32-11.72&***\\ 
edb&&$\checkmark$&&&251&15&332&133&94.47\%&93.18-95.85&28.63\%&26.95-30.59&33.06\%&30.36-36.72&***\\ 
ekits&&$\checkmark$&&&76&2&14&6&97.44\%&97.44-97.44&30.00\%&30-30&59.44\%&59.44-59.44&***\\ 
edbNOekits&&$\checkmark$&&&219&15&341&134&93.78\%&92.24-95.58&28.14\%&26.35-29.81&29.28\%&25.67-33.07&***\\ 
nvdNOekitsNOedb&&$\checkmark$&&&85&18&786&307&82.35\%&75.51-88.61&28.03\%&26.45-29.52&4.20\%&0.95-7.09&*\\ 
\hline nvd&$\checkmark$&$\checkmark$&&&87&10&716.5&247&89.50\%&83.67-94.57&25.65\%&24.48-26.95&6.87\%&3.87-9.67&***\\ 
edb&$\checkmark$&$\checkmark$&&&134&2&165&74&98.53\%&98.45-98.59&30.92\%&29.13-32.75&42.18\%&40.34-43.98&***\\ 
ekits&$\checkmark$&$\checkmark$&&&40&0&10&3&100.00\%&100-100&23.08\%&23.08-23.08&80.00\%&80-80&*\\ 
edbNOekits&$\checkmark$&$\checkmark$&&&106&2&166&74&98.15\%&98.04-98.23&30.83\%&29.1-32.53&36.34\%&34.52-37.67&***\\ 
nvdNOekitsNOedb&$\checkmark$&$\checkmark$&&&50&10&734&249&83.33\%&74.99-91.24&25.32\%&24.14-26.49&2.55\%&-0.09-4.89&\\ 
\hline nvd&&&$\checkmark$&&62&12&881&290&83.54\%&76.46-91.25&24.72\%&23.4-26.13&2.53\%&0.17-5.17&\\ 
edb&&&$\checkmark$&&127&5&634&157&96.34\%&94.07-98.55&19.81\%&18.58-21.03&13.81\%&11-16.62&***\\ 
ekits&&&$\checkmark$&&76&2&17&7&97.44\%&97.4-97.44&29.17\%&29.17-30.43&59.50\%&59.3-60&***\\ 
edbNOekits&&&$\checkmark$&&111&5&646&156&96.06\%&93.33-98.28&19.51\%&18.32-20.78&11.87\%&9.11-14.29&***\\ 
nvdNOekitsNOedb&&&$\checkmark$&&46&13&881&298&78.33\%&69.49-87.28&25.27\%&23.98-26.64&0.86\%&-1.27-3.06&\\ 
\hline nvd&$\checkmark$&&$\checkmark$&&21&2&285&48&91.30\%&77.78-100&14.50\%&13.17-15.92&2.78\%&-3.86-8.55&\\ 
edb&$\checkmark$&&$\checkmark$&&50&0&211&42&100.00\%&97.92-100&16.67\%&15.06-18.18&18.93\%&14.73-22.85&***\\ 
ekits&$\checkmark$&&$\checkmark$&&40&0&10&3&100.00\%&100-100&23.08\%&23.08-25&80.00\%&79.59-81.63&*\\ 
edbNOekits&$\checkmark$&&$\checkmark$&&41&0&218&42&100.00\%&97.43-100&16.23\%&14.84-17.69&15.46\%&11.95-19.07&**\\ 
nvdNOekitsNOedb&$\checkmark$&&$\checkmark$&&13&2&288.5&50&85.71\%&64.29-100&14.72\%&13.29-16.22&0.16\%&-6.2-4.99&\\ 
\hline nvd&&$\checkmark$&$\checkmark$&&153&31&719&288&83.14\%&78.36-87.64&28.57\%&26.99-30.12&7.77\%&4.34-11.1&***\\ 
edb&&$\checkmark$&$\checkmark$&&245&15&291&94&94.40\%&93.2-95.87&24.42\%&22.25-26.43&32.30\%&28.24-37.09&***\\ 
ekits&&$\checkmark$&$\checkmark$&&76&2&14&5&97.44\%&97.4-97.44&26.32\%&26.32-27.78&55.87\%&55.7-56.66&**\\ 
edbNOekits&&$\checkmark$&$\checkmark$&&215&15&300&94&93.69\%&92.14-95.07&23.88\%&21.88-25.71&28.43\%&24.62-32.12&***\\ 
nvdNOekitsNOedb&&$\checkmark$&$\checkmark$&&108&32&744&295&77.29\%&71.64-82.86&28.38\%&26.83-29.91&2.91\%&-0.23-6.04&\\ 
\hline nvd&$\checkmark$&$\checkmark$&$\checkmark$&&42&6&249&46&88.24\%&80.85-95.75&15.66\%&13.79-17.55&3.36\%&-4.99-11.57&\\ 
edb&$\checkmark$&$\checkmark$&$\checkmark$&&93&2&104&25&97.92\%&97.67-99&19.20\%&16.42-22.05&40.11\%&35.67-45.89&***\\ 
ekits&$\checkmark$&$\checkmark$&$\checkmark$&&40&0&10&3&100.00\%&100-100&23.08\%&23.08-25&80.00\%&79.59-81.63&*\\ 
edbNOekits&$\checkmark$&$\checkmark$&$\checkmark$&&76&2&109&25&97.50\%&97.1-98.77&18.32\%&16.03-21.05&34.15\%&29.7-39.06&***\\ 
nvdNOekitsNOedb&$\checkmark$&$\checkmark$&$\checkmark$&&25&6&262&47&81.82\%&66.67-92.87&15.21\%&13.31-17.04&-1.96\%&-11.33-5.62&\\ 
\hline nvd&&&&$\checkmark$&400&130.5&522&150&75.42\%&73.45-77.06&22.32\%&20.32-24.15&-3.13\%&-7.81-0.91&\\ 
edb&&&&$\checkmark$&288&19&175&40&93.87\%&93.07-94.9&18.78\%&16.74-20.8&30.59\%&25.83-36.56&***\\ 
ekits&&&&$\checkmark$&71&2&10&1&97.26\%&97.1-97.37&9.09\%&7.69-11.11&20.83\%&18.33-23.21&\\ 
edbNOekits&&&&$\checkmark$&246&19&179&41&92.91\%&92.07-93.94&18.50\%&16.66-20.39&26.35\%&21.66-31.28&***\\ 
nvdNOekitsNOedb&&&&$\checkmark$&265&131&546&155&67.01\%&65.05-69.23&22.14\%&20.14-24.02&-13.02\%&-17.15--8.73&***\\ 
\hline nvd&$\checkmark$&&&$\checkmark$&124&29&277&27&81.21\%&77.77-84.11&8.99\%&7.21-10.7&-20.45\%&-30.17--11.51&**\\ 
edb&$\checkmark$&&&$\checkmark$&124&2&61&4.5&98.41\%&98.32-98.5&6.90\%&5.63-8.33&35.90\%&31.07-41.14&\\ 
ekits&$\checkmark$&&&$\checkmark$&37&0&8&0&100.00\%&100-100&0.00\%&NA&NA&NA&\\ 
edbNOekits&$\checkmark$&&&$\checkmark$&98&2&62&5&98.00\%&97.89-98.1&7.25\%&5.71-8.2&31.55\%&25.41-35.63&\\ 
nvdNOekitsNOedb&$\checkmark$&&&$\checkmark$&67&28&296&28&70.30\%&64.13-74.51&8.61\%&6.71-10.26&-31.76\%&-41.83--23.8&***\\ 
\hline nvd&&$\checkmark$&&$\checkmark$&400&131&520&150&75.33\%&73.6-77.03&22.27\%&20.35-24.28&-3.29\%&-7.67-1.35&\\ 
edb&&$\checkmark$&&$\checkmark$&288&19&175&40&93.92\%&93.05-94.77&18.78\%&16.67-20.74&30.64\%&25.5-35.38&***\\ 
ekits&&$\checkmark$&&$\checkmark$&70&2&11&1&97.22\%&97.1-97.37&8.33\%&7.69-11.11&20.24\%&18.14-22.54&\\ 
edbNOekits&&$\checkmark$&&$\checkmark$&247&19&178&41&92.91\%&91.92-94.01&18.55\%&16.74-20.72&26.41\%&20.92-31.43&***\\ 
nvdNOekitsNOedb&&$\checkmark$&&$\checkmark$&265&131&548&155&66.92\%&64.56-69.14&22.13\%&20.12-23.99&-13.19\%&-17.82--8.82&***\\ 
\hline nvd&$\checkmark$&$\checkmark$&&$\checkmark$&123&29&276&27&80.95\%&78.18-83.87&9.02\%&7.35-10.86&-20.34\%&-29.76--12.79&**\\ 
edb&$\checkmark$&$\checkmark$&&$\checkmark$&123&2&61&5&98.40\%&98.31-98.48&6.81\%&5.63-8.34&35.72\%&30.55-41.49&\\ 
ekits&$\checkmark$&$\checkmark$&&$\checkmark$&37&0&8&0&100.00\%&100-100&0.00\%&NA&NA&NA&\\ 
edbNOekits&$\checkmark$&$\checkmark$&&$\checkmark$&98&2&62&4&98.00\%&97.87-98.11&6.50\%&5.63-8.2&30.10\%&25.31-36.17&\\ 
nvdNOekitsNOedb&$\checkmark$&$\checkmark$&&$\checkmark$&68&29&296&28&70.05\%&65.05-74.75&8.67\%&6.89-10.48&-31.90\%&-41.18--23.12&***\\ 
\hline nvd&&&$\checkmark$&$\checkmark$&577&180&345&91&76.19\%&75.03-77.36&20.94\%&18.53-23.46&-3.66\%&-7.98-0.76&\\ 
edb&&&$\checkmark$&$\checkmark$&316&20&86&14&94.03\%&93.49-94.64&13.74\%&11.34-16.33&19.29\%&13.18-25.65&*\\ 
ekits&&&$\checkmark$&$\checkmark$&70&2&7&1&97.22\%&97.06-97.33&12.50\%&9.09-20&24.12\%&20.51-27.86&\\ 
edbNOekits&&&$\checkmark$&$\checkmark$&270&20&88&14&93.03\%&92.41-93.66&13.73\%&11.65-16.35&16.17\%&11.04-22.63&\\ 
nvdNOekitsNOedb&&&$\checkmark$&$\checkmark$&375&180.5&376&96&67.56\%&65.99-68.89&20.25\%&18.3-22.3&-15.29\%&-19.43--11.41&***\\ 
\hline nvd&$\checkmark$&&$\checkmark$&$\checkmark$&134&32&147&16&80.59\%&78.02-83.23&9.82\%&7.14-12.42&-19.10\%&-28.67--10.07&*\\ 
edb&$\checkmark$&&$\checkmark$&$\checkmark$&118&2&37&2&98.33\%&98.25-98.43&5.13\%&4.55-6.06&25.99\%&22.73-29.74&\\ 
ekits&$\checkmark$&&$\checkmark$&$\checkmark$&34&0&6&0&100.00\%&100-100&0.00\%&0-0&NA&NA&\\ 
edbNOekits&$\checkmark$&&$\checkmark$&$\checkmark$&96&2&39&2&97.96\%&97.83-98.06&4.88\%&4.44-5.56&21.22\%&17.91-24.63&\\ 
nvdNOekitsNOedb&$\checkmark$&&$\checkmark$&$\checkmark$&74&33&163&16&69.52\%&65.34-73.59&8.94\%&6.39-11.56&-35.79\%&-45.5--26.2&***\\ 
\hline nvd&&$\checkmark$&$\checkmark$&$\checkmark$&578&180&345&91&76.22\%&75.06-77.39&20.86\%&18.62-23.27&-3.85\%&-8.02-0.31&\\ 
edb&&$\checkmark$&$\checkmark$&$\checkmark$&316&20&86&14&94.05\%&93.47-94.64&13.99\%&11.34-16.5&19.40\%&13.26-25.78&*\\ 
ekits&&$\checkmark$&$\checkmark$&$\checkmark$&70&2&7&1&97.22\%&97.06-97.33&12.50\%&10-16.67&24.06\%&21.17-26.92&\\ 
edbNOekits&&$\checkmark$&$\checkmark$&$\checkmark$&270&20&87&14&92.98\%&92.36-93.64&13.86\%&11.65-16.16&16.32\%&10.48-21.76&*\\ 
nvdNOekitsNOedb&&$\checkmark$&$\checkmark$&$\checkmark$&376&180&376&96&67.62\%&65.94-69.12&20.40\%&18.06-22.74&-15.18\%&-19.57--10.75&***\\ 
\hline nvd&$\checkmark$&$\checkmark$&$\checkmark$&$\checkmark$&135&32&147&16&80.62\%&78.23-83.33&9.66\%&6.83-12.26&-19.37\%&-28.97--9.98&*\\ 
edb&$\checkmark$&$\checkmark$&$\checkmark$&$\checkmark$&119&2&37&2&98.35\%&98.25-98.43&5.13\%&4.55-5.88&25.97\%&22.72-29.23&\\ 
ekits&$\checkmark$&$\checkmark$&$\checkmark$&$\checkmark$&35&0&6&0&100.00\%&100-100&0.00\%&0-0&NA&NA&\\ 
edbNOekits&$\checkmark$&$\checkmark$&$\checkmark$&$\checkmark$&96&2&39&2&97.96\%&97.85-98.08&4.88\%&4.44-5.56&21.32\%&17.94-24.44&\\ 
nvdNOekitsNOedb&$\checkmark$&$\checkmark$&$\checkmark$&$\checkmark$&74&32&163&16&69.44\%&65.04-73.69&8.81\%&6.59-11.17&-36.00\%&-46.05--26.12&***\\\hline
\end{tabular}
\end{table*}

\end{document}